\documentclass[proof]{WileyASNA-v1}

\articletype{Article Type}%

\received{1 October 2020}
\revised{12 October 2020}
\accepted{25 October 2020}

\usepackage{lipsum}
\usepackage{comment}

\usepackage{float}

\def\bea{\begin{eqnarray}}
\def\beann{\begin{eqnarray*}}
\def\eea{\end{eqnarray}}
\def\eeann{\end{eqnarray*}}

\def\nn{\nonumber}

\def\hst{\hspace{.5cm};\hspace{.5cm}}

\begin{document}

\title{Ground Level Muon Flux Variation in a Cosmic Rays Simulation}

\author[1]{E. G. Pereira, C. A. Z. Vasconcellos, D. Hadjimichef\,*}

\authormark{Elisa Garcia Pereira}

\address[1]{\orgdiv{Instituto de F\'isica}, \orgname{Universidade Federal do Rio Grande do Sul (UFRGS)}, \orgaddress{\state{Porto Alegre}, \country{Brazil}}}


\corres{*Av. Bento Gon\c{c}alves, 9500 - Agronomia, Porto Alegre - RS, 91501-970.  \email{dimihadj@gmail.com}}


\abstract{
The oscillatory movements of   atmospheric air masses has been claimed to be the origin of the
muon flux variation  measured at the ground level. Using a cosmic ray toolkit (CORSIKA),
we simulate cascade scenarios in a   time scale of a   year and show the dependence of the muon flux pattern
with a proposed oscillatory model atmosphere.}
 

\maketitle


\section{Introduction}

High energy cosmic rays, mostly protons, but also alpha particles and heavy nucleus,  hit the air molecules starting an 
extensive air  shower.  A shower is a cascade of many kilometers of ionized particles and electromagnetic radiation. 
As the fragments of the first collision hit other nuclei, a  cascade of pions is produced. 
The neutral pions initiate an electromagnetic shower constituted
of photons, electrons and positrons. The charged pions
will interact with other atoms or in turn, decay into muons and neutrinos,
constituting the most prominent flux at the ground level.

The role played by the atmosphere is that of a giant calorimeter, where the ground level flux is stongly dependent 
on the atmospheric density. There are several mechanisms that lead to changes in atmospheric density.
For example, solar heating of
atmospheric layers is the dominant mechanism, that gives origin to atmospheric tides
\citep{lindzen1,lindzen2,lindzen3,lindgren}. These
tides are oscillatory movements of air masses characterized by a set of accurately known
frequencies that reflect the amount of daily insolation as Earth revolves around the Sun.  
In  principle these air density  tides  could effect the muon flux at the ground level.
In the experiment performed by Takai and collaborators they 
report  the detection of tidal frequencies in the spectral analysis of time series muon flux
measurements realized over a period of eight years   \citep{takai}.

An alternative approach to the direct measurement of cosmic rays has been the   development of powerful simulation tools, based on
Monte Carlo techniques that incorporate the complex physical content of the  extensive air 
shower. In particular, the  CORSIKA (COsmic Ray SImulations for KAscade) toolkit
 is a code  for detailed simulation of these showers that are initiated by high energy cosmic ray particles. 
It allows one to study the interactions and decay of nucleus, hadrons, muons, electrons, and photons of high energy, over $10^{20}$  eV, in the atmosphere \citep{corsika}.

In this work, we shall examine the periodic variation of the  atmospheric density  as an influence  on the muon flux counting at the ground level, using the CORSIKA toolkit  simulations. Our main objective is to compare the simulated results with experimental data that exhibits a tidal frequency behavior.

\section{Muon flux measurement}

In  this section we shall briefly review the muon detection experiment of reference \citep{takai}, that
  reports   the detection of tidal frequencies in the spectral analysis of time series muon flux
measurements realized over a period of eight years. The muon telescope used for these
measurements was part of the MARIACHI experiment,   located at Smithtown High School
East in the state of New York, latitude 40$^{\circ}$ 52' 14.88"N, longitude 73$^{\circ}$ 9' 53.103"W and 43 m above
sea level. The large-scale oscillations of the atmosphere, producing tides   are those, in general,
generated by (a) the gravitational forces of the moon and sun, and (b) the thermal action of the sun \citep{lindzen1}.

 The detection system consisted of two 0.28 m$^2$ plastic scintillators subtending a solid
angle of 3.8 sr. Counts per minute were recorded by a computer and assigned a time stamp provided
by a  GPS clock with a nominal accuracy of 100 ns. The setup was
located indoors with approximately 19 g/cm$^2$ of roofing material above the detectors. The muons were detected with
momenta above 200 MeV/c.  They recorded  a total
number of samples  of $3.391\times 10^6$. The average measured muon rate was (1890 $\pm$ 51) counts/min, which corresponds
 to a rate of 29.6  $\pm$ 0.8 counts/s\, m$^2$ sr. 

Figure  \ref{takai} is extracted from \citep{takai} where 
an hourly-averaged, pressure corrected time series of muon data is seen.
A striking feature is a yearly modulation with an amplitude of $\pm$5\% of the average
counts, with maxima and minima during winter and summer seasons, respectively. This
modulation is caused by seasonal variations in solar heating that expand or contract the
atmosphere. This change in atmospheric thickness alters, increasing or decreasing,
the muon flight path.

\section{The Numerical Model}

To simulate a similar effect, as the experiment described in the last section, we used  the CORSIKA  toolkit (version 7100),
 which is a detailed Monte Carlo
program designed to study the evolution of extensive air  showers  in the atmosphere initiated by photons, 
protons, nuclei, or any other particle. It was originally developed to perform simulations
for the KASCADE experiment \citep{kascade1,kascade2} at Karlsruhe and has been refined over the past
years. The program recognizes 50 elementary particles: $\gamma$, $e^{\pm}$, $\mu^{\pm}$,
$\pi^0$, $\pi^{\pm}$, $K^{\pm}$, $K^{0}_{\rm S/L}$, $\eta$; the baryons p, n, $\Lambda$, $\Sigma^{\pm}$, $\Sigma^0$, 
$\Xi^0$, $\Xi^{-}$, $\Omega^{-}$,   the corresponding anti-baryons, the resonance states $\rho^{\pm}$, 
 $\rho^{0}$,  $K^{\ast\,\pm}$, $K^{\ast\,0}$, $\bar{K}^{\ast\,0}$, $\Delta^{++}$, $\Delta^{+}$, $\Delta^{0}$, $\Delta^{-}$
 and the
corresponding anti-baryonic resonances. Optionally the neutrinos $\nu_e$  and $\nu_\mu$  and  anti-neutrinos 
 $\bar{\nu}_e$  and $\bar{\nu}_\mu$ resulting from $\pi$, $K$, and $\mu$ decay may be generated explicitly. In
addition nuclei up to A = 56 can also be treated.

\begin{figure}[h!]
\centering
\includegraphics[width=0.8\linewidth]{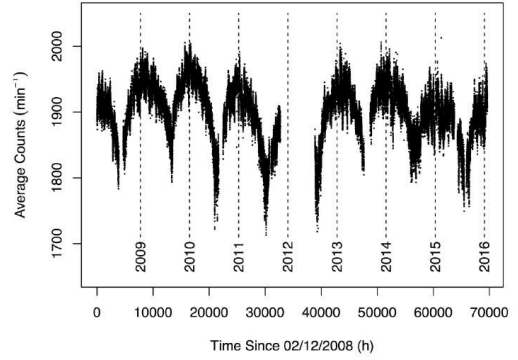} 
\caption{Measurement  of muon counts over a period of 8 years from  \citep{takai}.} 
\label{takai}
\end{figure}

To adapt  CORSIKA's code for our problem, first we   changed its execution parameters in an automated way, 
by   creating simple auxiliary codes in bash, Python 3, C, and R.
We assume that  the primary particle is always a proton with an energy of 10 TeV. In order to compare with  Fig. \ref{takai}, 
our simulation is set for  New York,   choosing  the local magnetic field \citep{magnet} and   placing  the detector $43$ m above sea level, as is described in \citep{takai}. Our simulation is for a time span of one year.
To improve the numerical performance, the seed of each run is random and the final data is the averaged
 over five runs with the same parameters but different seeds.  The atmospheric data was extracted from 
CORSIKA's documentation and the magnetic field  parameters   from the NOAA's website \citep{magnet}.

A crucial step in our calculation is the definition of the atmospheric model.
CORSIKA adopts the Earth's atmospheric composition as $78.1\%$ N$_2 $, $21.0\%$ O$_2$, $0.9\%$  Ar and its density variation  is modeled by 5 layers. In the four lower layers, the density function $T(h)$ has an exponential dependence with the height $h$, while a linear dependence in the fifth layer:
\begin{subequations}
\bea
\hspace{-3cm}
T(h) &=&   a_i   +  b_i\,e^{-h/c_i}
\label{corsika-atm1}
\\ 
T(h) &=& a_5 - b_5 \cdot h/c_5\,\,.
\label{corsika-atm2}
\eea
\end{subequations}
All layers are parameterized by coefficients $a_i$, $b_i$ and $c_i$ with $i=1,\ldots, 5$, defined in table \ref{tab:atm},
where we adopt the U.S. standard atmosphere,   as is presented in the CORSIKA documentation.

In order to incorporate the oscillatory movements of   atmospheric air masses
we must modify   the density function $T(h)\to T(h,t)$, with the inclusion of a periodic time dependence in the four lower layers by the
following substitution
\bea
a_i &\to& a_i(t)=a_i + f(t) \hst b_i \to b_i(t)=b_i + f(t) 
\nn\\
c_i &\to& c_i(t)=c_i + f(t)
\label{model-par}
\eea
with $i =1,\ldots, 4$. We have chosen the same time dependent function $f(t)$ for all the coefficients 
and assume   
\bea
f(t)= B \sin^2(\omega\,t)\,,
\label{atm3}
\eea
where $\omega$ is $\pi/364$,\, $t$ is measured in days and $B$ is a free parameter.  
The original CORSIKA atmosphere is regained by setting  the phenomenological parameter $B=0$ in (\ref{atm3}).
A plot of the new $T(h,t)$ is seen in Fig \ref{th} with four different $B$ values and at a  height of $h=6$ km.

\begin{table}
\begin{center}\vspace{1cm}
\begin{tabular}{c c c c c }
\toprule
$i$ & $h$ (km) & $a_i$  (g/cm$^2$)  & $ b_i$  (g/cm$^2$)   & $c_i$  (g/cm$^2$) \\
\midrule
1 &	0-4 & -186.555305 & 1222.6562 &	994186.38\\
2 &	4-10 &	-94.919 & 1144.9069 & 878153.55\\
3 & 10-40 & 0.61289 & 1305.5948 & 636143.04\\
4 & 40-100 & 0 & 540.1778 &	772170.16\\
5 &	$>$100 &	0.01128292 &	1 &	10000000000\\
\bottomrule
\end{tabular}
\captionof{table}{Parameters of the U.S. standard atmosphere  }
\label{tab:atm}
\end{center}\vspace{1cm} 
\vfill\null \columnbreak
\end{table}
\section*{Results}

\begin{figure}[!htbp]
\begin{center}
\includegraphics*[width=1.0\linewidth]{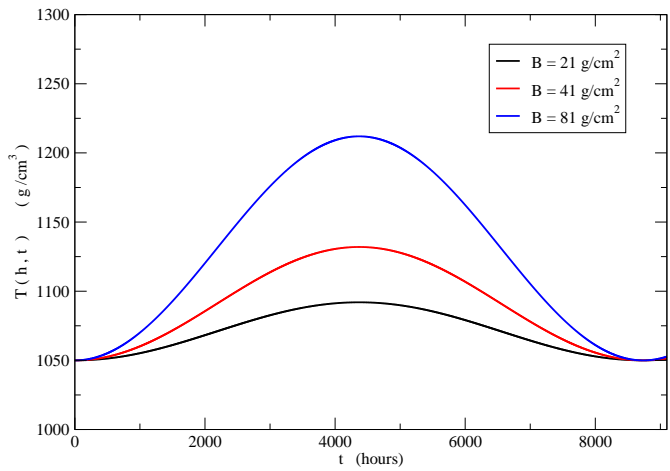}
\captionof{figure}{The model atmospheric density function $T(h,t)$}
\label{th}
\end{center}
\end{figure}

\begin{figure}[!htbp]
\begin{center}
\includegraphics*[width=1.0\linewidth]{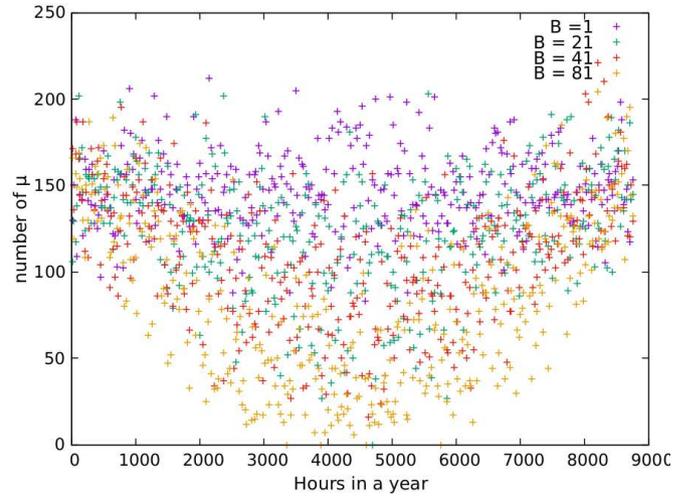}
\captionof{figure}{Muon counts over a period of a year, for different $B$ values}
\label{img:sin}
\end{center}
\end{figure}

In Fig. \ref{img:sin} is shown  the simulation  for the  number of muons per day in a year for different $B$ values.
The behavior is consistent with the results presented in Fig \ref{th}, where an increase in $B$   implies a decrease 
in the number of muons, in the middle of the year (when $\sin^2(\omega t)$ is maximum). These results are in agreement 
with an annual   measurement 
seen in Fig. \ref{takai}


 \begin{figure}[!htbp]
\centering
\includegraphics*[width=1\linewidth, height=9cm]{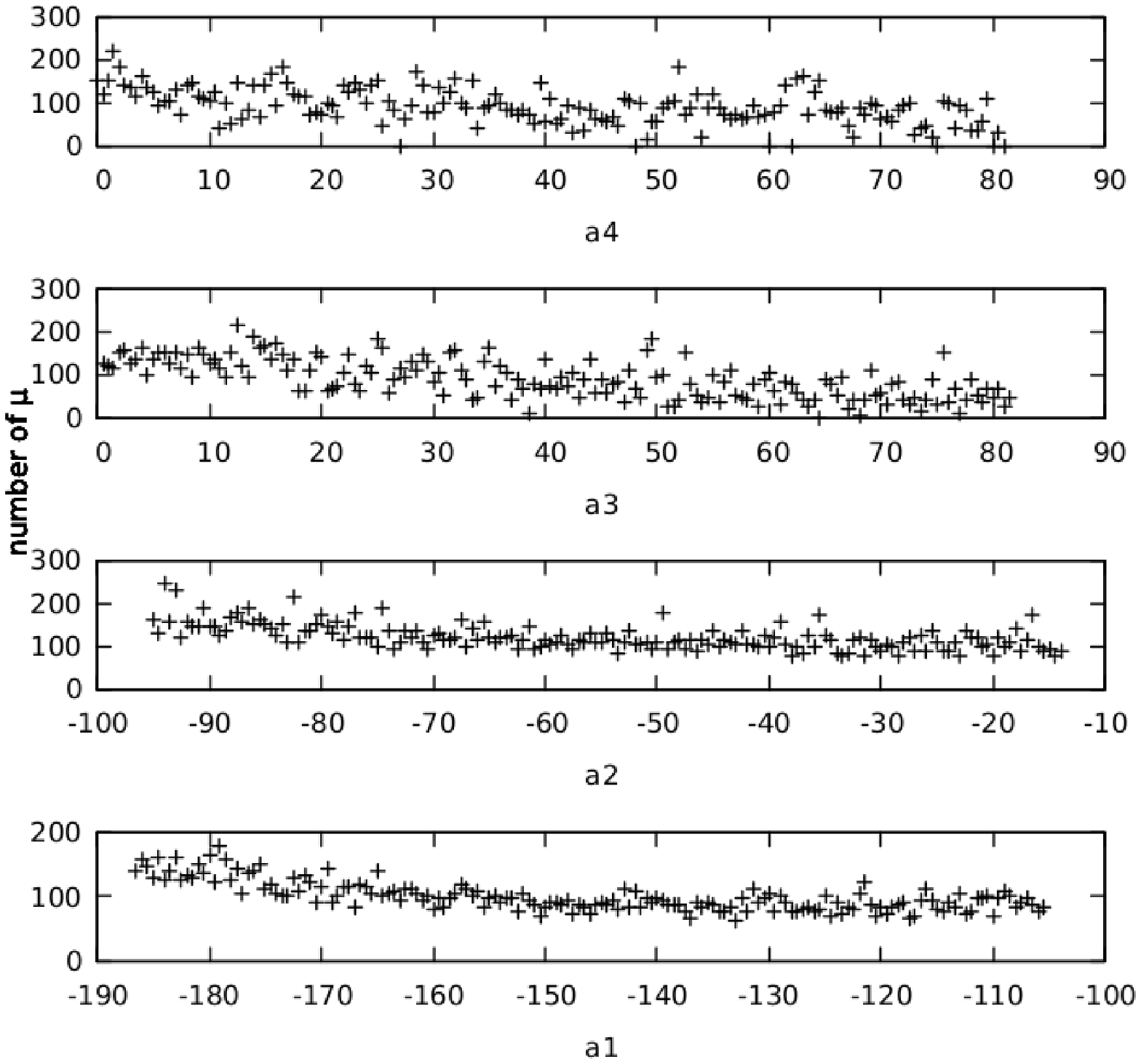}
\caption{$a_i$ for $i=1,2,3,4$ for $B=81$ g/cm$^2$} 
\label{plot}
\end{figure}

 \begin{figure}[!htbp]
\centering
\includegraphics*[width=1\linewidth, height=9cm]{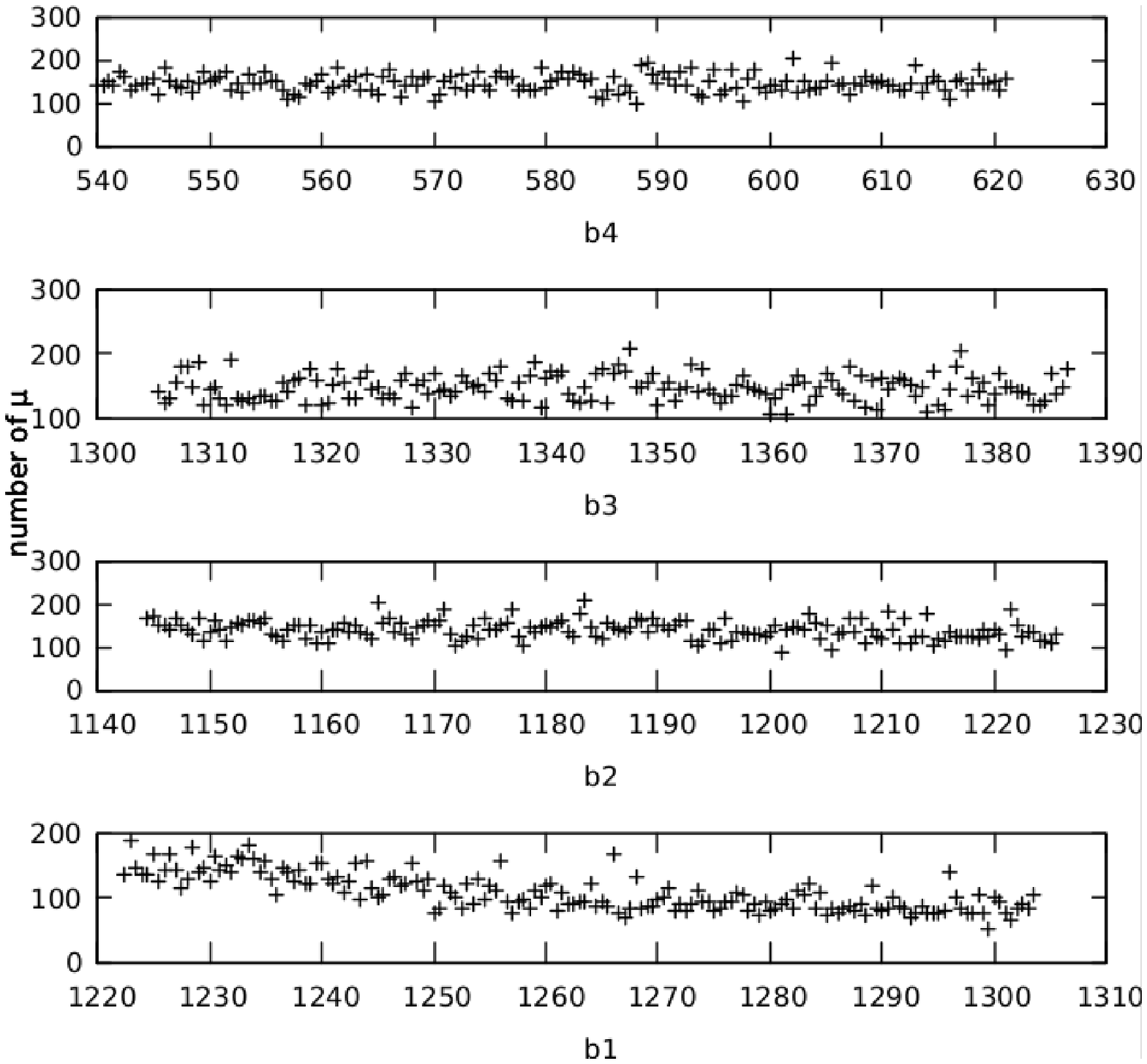}
\caption{$b_i$ for $i=1,2,3,4$ for $B=81$ g/cm$^2$}
\label{plot2}
\end{figure}

 \begin{figure}[!htbp]
\centering
\includegraphics*[width=1\linewidth, height=9cm]{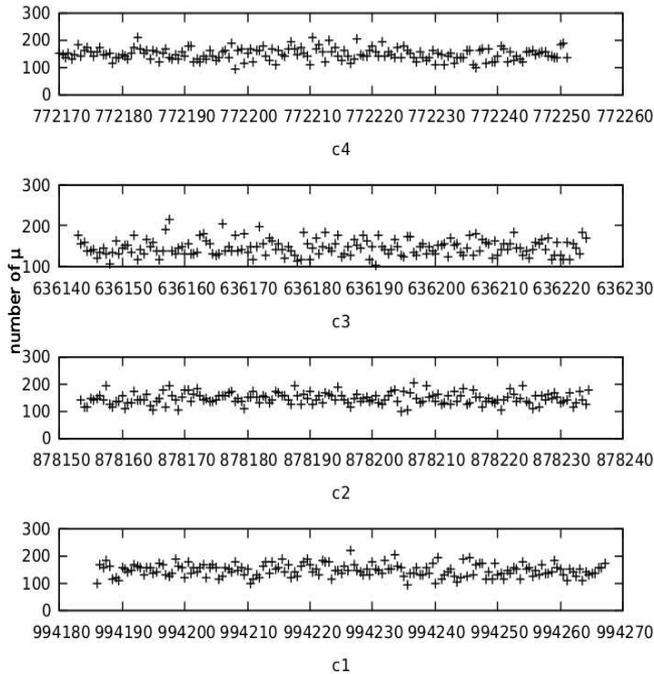}
\caption{$c_i$ for $i=1,2,3,4$ for $B=81$ g/cm$^2$}
\label{plot3}
\end{figure}

To test the sensibility of this effect in each atmospheric layer, figures 
\ref{plot}- \ref{plot3} are plots of muon counts while varing parameters 
$a_i$, $b_i$ and  $c_i$. The strategy is to keep  all the other parameters     constant, according to table \ref{tab:atm}, 
as we change, with a simple linear variation,   each parameter individually. 
For the sake of comparison we choose the case $B=81$ g/cm$^2$ in Figures \ref{th} and \ref{img:sin} .
Therefore each parameter starts from the original tabulated value,  increased   with a step of 0.5 g/cm$^2$,  stopping when the  sum of the 
increments add up to  81 g/cm$^2$.
In the lowst atmospheric layer ($i=1$), from ground level up to 4 km, more muons are detected for smaller parameter values,
as seen  in  figures \ref{plot} and \ref{plot2} for  $a_1$ and $b_1$, respectively. 
Which means that in the layer closer to the ground level, the $\sin^2(wt)$ weight is more influential.
For the other $a_i$ values we still have a slow and subtle decrease in the number of muons detected on the ground level.
As for the other $b_i$ values no significant change in the number of muons is present for the upper atmospheric layers.
 The same conclusion is valid for all the $c_i$ values calculated.


\section{Conclusions}

Notwithstanding the simulation was restricted to the range of a year,
due to computation time, it is clear   that, for the chosen atmospheric density range, the lower layers are more influential 
 on the number of muons. 
We have also shown that a simple  phenomenological periodic time-dependent   density function $T(h,t)$ can reproduce qualitatively
the complex atmospheric tides effects that are revealed in the muon data.

We aim in our forthcoming research to explore more realistic   atmospheric models, 
based on the theory of atmospheric   and thermal tides.  This  investigation can provide the sources of periodic excitation
 and  atmospheric response to the excitation.

\bibliography{elisa_iwara2020}%

\section*{Acknowledgements}

The authors thank  the  High Energy Physics Simulations  group (HEPsim, 
\texttt{www.ufrgs.br/hepsim})
  for all the help and discussion in high energy physics simulations.

\end{document}